\documentclass[graybox, secnum]{svmult}

\usepackage{mathptmx}       
\usepackage{helvet}         
\usepackage{courier}        
\usepackage{type1cm}        
\usepackage{makeidx}         
\usepackage{graphicx,psfrag,epsfig}        
\usepackage{multicol}        
\usepackage{multirow}
\usepackage[bottom]{footmisc}
\usepackage{hyperref}        
\usepackage{soul}            
\hypersetup{colorlinks=true,urlcolor=blue}
\usepackage[numbers,sort&compress]{natbib}
\makeindex             

\def\prd{Phys. Rev. D}
\def\prl{Phys. Rev. Lett.}
\def\aap{Astron. Astroph.}
\def\apj{Astrophys. J.}
\def\apjl{Astrophys. J. Lett.}
\def\apjs{Astrophys. J. Suppl.}
\def\mnras{Mon. Not. R. Astron. Soc.}
\def\nar{New Astron. Rev.}
\def\nat{Nature}
\def\jcap{JCAP}

\begin{document}
\title*{Tests of Lorentz Invariance}
\author{Jun-Jie Wei \thanks{corresponding author} and Xue-Feng Wu \thanks{corresponding author}}
\institute{Jun-Jie Wei \at Purple Mountain Observatory, Chinese Academy of Sciences, Nanjing 210023, China
\at School of Astronomy and Space Sciences, University of Science and Technology of China, Hefei 230026, China \\ \email{jjwei@pmo.ac.cn}
\and Xue-Feng Wu \at Purple Mountain Observatory, Chinese Academy of Sciences, Nanjing 210023, China
\at School of Astronomy and Space Sciences, University of Science and Technology of China, Hefei 230026, China \\ \email{xfwu@pmo.ac.cn}}
%
%
\maketitle
\abstract{Lorentz invariance is a fundamental symmetry of both Einstein's theory of general relativity
and quantum field theory. However, deviations from Lorentz invariance at energies approaching the Planck scale
are predicted in many quantum gravity theories seeking to unify the force of gravity with the other
three fundamental forces of matter. Even though any violations of Lorentz invariance are expected to be
very small at observable energies, they can increase with energy and accumulate to detectable levels over
large distances. Astrophysical observations involving high-energy emissions and long baselines can therefore
offer exceptionally sensitive tests of Lorentz invariance. With the extreme features of astrophysical
phenomena, it is possible to effectively search for signatures of Lorentz invariance violations (LIV)
in the photon sector, such as vacuum dispersion, vacuum birefringence, photon decay, and photon splitting.
All of these potential signatures have been studied carefully using different methods over the past few decades.
This chapter attempts to review the status of our current knowledge and understanding of LIV, with particular emphasis
on a summary of various astrophysical tests that have been used to set lower limits on the LIV energy scales.}

\keywords{Lorentz invariance, Astroparticle physics, Quantum gravity, Relativity and gravitation}

\section{Introduction}
Einstein's theory of general relativity gives a classical description of gravity,
and the Standard Model of particle physics is a well-tested quantum theoretical model of particles and
all fundamental forces except gravity. They provide an excellent description of nature together at experimentally
attainable energies. However, at the Planck scale ($E_{\rm Pl}=\sqrt{\hbar c^{5}/G}\simeq1.22\times10^{19}$ GeV),
a more fundamental quantum theory of gravity is required, which would be a giant leap towards unification
of all fundamental forces. A correct and consistent unification scheme has long been the Holy Grail
of modern physics. Several quantum gravity (QG) theories that attempt to unify general relativity and
the standard Model at the Planck scale predict that Lorentz invariance may be broken at this energy scale,
perhaps due to the underlying quantized nature of spacetime \cite{1989PhRvD..39..683K,1991NuPhB.359..545K,
1995PhRvD..51.3923K,2005LRR.....8....5M,2005hep.ph....6054B,2013LRR....16....5A,2014RPPh...77f2901T,2021FrPhy..1644300W}.
Therefore, the dedicated experimental searches for violations of Lorentz invariance may help to clear the path
to a unification theory. Such experimental tests obtained from a wide range of systems have been compiled in Ref.~\cite{2011RvMP...83...11K}.

Although any deviations from Lorentz invariance are supposed to be very tiny at attainable energies $\ll E_{\rm Pl}$,
they can increase with energy and over large distances due to cumulative process in particle propagation.
Astrophysical observations involving high energies and long baselines can therefore provide the most sensitive tests of
Lorentz invariance. In the photon sector, the potential signatures of Lorentz invariance violations (LIV) include
vacuum dispersion, vacuum birefringence, photon decay, photon splitting, and so on \cite{2008ApJ...689L...1K,
2001APh....16...97S}.

Vacuum dispersion would produce an energy-dependent velocity of light, which would lead to
arrival-time differences between promptly emitted photons with different energies traveling over astrophysical distances.
Astrophysical time-of-flight measurements can therefore be used to test Lorentz invariance
(e.g., Refs.
\cite{1998Natur.393..763A,
2006APh....25..402E,
2008JCAP...01..031J,
2008ApJ...689L...1K,
2009PhRvD..80a5020K,
2009Sci...323.1688A,
2009Natur.462..331A,
2012APh....36...47C,
2012PhRvL.108w1103N,
2013PhRvD..87l2001V,
2013APh....43...50E,
2015PhRvD..92d5016K,
2015APh....61..108Z,
2017ApJ...834L..13W,
2017ApJ...842..115W,
2017ApJ...851..127W,
2019PhRvD..99h3009E,
PhysRevLett.125.021301,
1999PhRvL..83.2108B,
Kaaret1999}).
Similarly, the effects of vacuum birefringence would accumulate over cosmological distances resulting in
a measurable rotation of the polarization plane of linearly polarized photons as a function of energy.
Thus, Lorentz invariance can also be tested with astrophysical polarization measurements (e.g., Refs.
\cite{1990PhRvD..41.1231C,
1998PhRvD..58k6002C,
2001PhRvD..64h3007G,
2001PhRvL..87y1304K,
2006PhRvL..97n0401K,
2007PhRvL..99a1601K,
2013PhRvL.110t1601K,
2003Natur.426Q.139M,
2004PhRvL..93b1101J,
2007MNRAS.376.1857F,
2009JCAP...08..021G,
2011PhRvD..83l1301L,
2011APh....35...95S,
2012PhRvL.109x1104T,
2013MNRAS.431.3550G,
2014MNRAS.444.2776G,
2016MNRAS.463..375L,
2017PhRvD..95h3013K,
2019PhRvD..99c5045F,
2019MNRAS.485.2401W}).
Since time-of-flight measurements are less sensitive than polarization measurements by a factor $\propto1/E$,
where $E$ is the energy of the light, time-of-flight measurements typically place less stringent limits on LIV
\cite{2009PhRvD..80a5020K}. However, numerous predicted signals of LIV have no vacuum birefringence, so
constraints from time-of-flight measurements are indispensable in a broad-based search for nonbirefringent
Lorentz-violating effects.

Besides the measurements of time-of-flight and polarization, astrophysical spectrum measurements can also be
used to test Lorentz invariance (e.g., Refs.
\cite{2001APh....16...97S,
2017PhRvD..95f3001M,
2019JCAP...04..054A,
2019EPJC...79.1011S,
2020PhRvL.124m1101A,
2021arXiv210507927C,
2021arXiv210612350T}).
This is because superluminal LIV allows photons to decay at high energies. Photons decay into electron-positron
pairs proceed over very short distances (centimeters or less) once above the energy threshold of the decay process,
which would result in a hard cutoff in the gamma-ray spectra of astrophysical sources \cite{2017PhRvD..95f3001M}.
Another superluminal LIV decay process, photon splitting into multiple photons, also predicts a cutoff at the highest
energy part of the photon spectra of astrophysical sources \cite{2019JCAP...04..054A,2019EPJC...79.1011S}.
Experimental searches for LIV-induced spectral cutoff have therefore been carried out in the observed spectra of
several astrophysical sources \cite{2017PhRvD..95f3001M,2019JCAP...04..054A,2019EPJC...79.1011S,2020PhRvL.124m1101A,
2021arXiv210507927C,2021arXiv210612350T}.

In this chapter, we summarize the current status on astrophysical tests of Lorentz invariance and attempt to chart
the future of the subject. We start by reviewing the recent achievements in sensitivity of vacuum dispersion
time-of-flight measurements. After that, we overview the progresses on LIV tests through the effects of vacuum birefringence,
photon decay, and photon splitting. Finally, we present a summary and future prospect.

\section{Vacuum dispersion}

\subsection{Modified photon dispersion relation}
Compared to the standard energy-momentum relationship in Einstein's special relativity, the introduction of LIV can
induce modifications to the photon dispersion relation in vacuum, which can be described using
a Taylor series \cite{1998Natur.393..763A}:
\begin{equation}
E^{2}\simeq p^{2}c^{2}\left(1\pm\sum_{n=1}^{\infty}a_{n}\right)\;,
\label{eq:LIVdispersion}
\end{equation}
where $E$ and $p$ are the energy and momentum of the photon, respectively, $c$ is the Lorentz-invariant speed of light, and
$a_{n}=(E/E_{{\rm QG},n})^{n}$ is a dimensionless expansion coefficient. Also, $E_{{\rm QG},n}$ is
the hypothetical energy scale at which QG effects become significant. At small energies $E\ll E_{{\rm QG},n}$, the sum is
dominated by the lowest-order term of the series. Considering the sensitivity of current detectors, only the first two leading
terms ($n=1$ or $n=2$) are of interest for independently experimental searches. They are often referred to as linear
and quadratic energy-dependent modifications, respectively. The energy-dependent photon group velocity can therefore
be derived as
\begin{equation}
\upsilon(E)=\frac{\partial E}{\partial p}\approx c\left[1\pm\frac{1+n}{2}\left(\frac{E}{E_{{\rm QG},n}}\right)^{n}\right]\;,
\label{eq:vLIV}
\end{equation}
where the sign $\pm$ allows for the ``superluminal'' ($+$) or ``subluminal'' ($-$) scenarios.

Owing to the energy dependence of $\upsilon(E)$, two photons with different energies radiated simultaneously from the same source would
travel at different speeds, thus arriving at the observer at different times. For example, in the subluminal case (where high-energy
photons are slower), the photon with a higher energy (denoted by $E_{h}$) would arrive with respect to the photon with a lower
energy ($E_{l}$) by a time delay
\cite{2008JCAP...01..031J}
\begin{equation}
\Delta t_{\rm LIV}=t_{h}-t_{l}
                  =\frac{(1+n)}{2H_{0}}\frac{E_{h}^{n}-E_{l}^{n}}{E_{{\rm QG}, n}^{n}}
\int_{0}^{z}\frac{(1+z')^{n}dz'}{\sqrt{\Omega_{\rm m}(1+z')^{3}+\Omega_{\Lambda}}}dz'\;,
\label{eq:tLIV}
\end{equation}
where $t_{h}$ and $t_{l}$ represent the observed arrival times of the high- and low-energy photons, respectively, and
$z$ is the redshift of the source. Here $H_{0}$, $\Omega_{\rm m}$, and $\Omega_{\Lambda}$ are cosmological parameters
of the standard $\Lambda$CDM model.

In principle, Lorentz invariance can be tested using the observed time delays of photons with different energies arising in
astronomical sources. However, it is difficult to exclude the possibility that the energy-dependent delays are of pure
astrophysical origins. In order to prove the existence of LIV, one needs to systematically collect numerous delay events
and demonstrate that they all match the prediction of LIV (Eq.~(\ref{eq:tLIV})).

On the contrary, it is relatively simple to place limits on the non-existence of LIV at a certain energy scale.
A smaller-than-expected observed delay of high-energy photons with respect to low-energy photons can be used to discard
a given model. More specifically, a conservative lower limit on the QG energy scale $E_{\rm QG}$ can be placed
under the assumption that the observed time delay is mainly contributed by the LIV effects.

\subsection{Present constraints from time-of-flight measurements}
It is obvious from Eq.~(\ref{eq:tLIV}) that a potential LIV-induced time delay is predicted to be an increasing function
of the photon energy and the source distance. Additionally, a short timescale of the signal variability provides a natural reference time
to measure the energy-dependent delay more easily. The greatest sensitivities on $E_{{\rm QG},n}$ can therefore be expected from those
violent astrophysical phenomena with high-energy emissions, large distances, and short time delays. The first attempt to
constrain LIV by exploiting the rapid variations of gamma-ray emissions from astrophysical sources at cosmological distances
was presented in Ref.~\cite{1998Natur.393..763A}. As of now, transient or variable sources such as gamma-ray bursts (GRBs),
active galactic nuclei (AGNs), and pulsars form a group of excellent candidates for searching for the LIV-induced
vacuum dispersion.

\subsubsection{Gamma-ray bursts}
GRBs are among the most distant gamma-ray transients in the Universe and their prompt emission light curves vary on
subsecond timescales. As such, GRBs are ideal probes that one can use to perform LIV tests. There have been lots of
LIV studies by applying various analysis techniques on the time-of-flight measurements of GRBs.

Some of the pre-Fermi
constraints are those by Ref.~\cite{Ellis2003} using BATSE GRBs; by Ref.~\cite{2004ApJ...611L..77B} using RHESSI
observations of GRB 021206; by Ref.~\cite{2008ApJ...676..532B} using HETE-2 GRBs; by Ref.~\cite{2008GReGr..40.1731L}
using INTEGRAL GRBs; by Ref.~\cite{2006JCAP...05..017R} using Konus-Wind and Swift observations of GRB 051221A; by
Refs.~\cite{2006APh....25..402E,2008APh....29..158E} using BATSE, HETE-2, and Swift GRBs; and by Ref.~\cite{2017ApJ...851..127W}
using Swift GRBs. Much more stringent constraints on LIV, however, have been obtained using Fermi observations,
entirely thanks to the unprecedented sensitivity for detecting GRB prompt MeV/GeV emission by the Large Area Telescope
(LAT) onboard the Fermi satellite. These tighter constraints include those by the Fermi Gamma-Ray Burst Monitor (GBM)
and LAT Collaborations using GRBs 080916C \cite{2009Sci...323.1688A} and 090510 \cite{2009Natur.462..331A};
by Ref.~\cite{2009PhRvD..80k6005X} using GRB 090510; and by Refs. \cite{2010APh....33..312S,2012APh....36...47C,2012PhRvL.108w1103N,2013PhRvD..87l2001V,2019PhRvD..99h3009E}
using multiple Fermi GRBs. Remarkably, Ref.~\cite{2009Natur.462..331A} studied the time-of-flight effect
by analyzing the arrival time lag between the highest energy (31 GeV) photon and low-energy photons from GRB 090510.
The burst was localized at $z=0.903$. Fermi/LAT detected this 31 GeV photon 0.829 s after the GBM trigger.
Assuming the 31 GeV photon was emitted at the GBM trigger time, the time delay between the highest energy photon
and low-energy (trigger) photons is then $\Delta t_{\rm LIV}<829$ ms. This gives the most conservative constraint on
the linear LIV energy scale $E_{{\rm QG}, 1}>1.19E_{\rm Pl}$. If the 31 GeV photon is assumed to be associated
with the contemporaneous $<1$ MeV spike, one therefore has $\Delta t_{\rm LIV}<10$ ms, which gives the most radical constraint
$E_{{\rm QG}, 1}>102E_{\rm Pl}$. These results obviously disfavor the linear LIV models requiring $E_{{\rm QG}, 1}\leq E_{\rm Pl}$,
and are in good agreement with Lorentz invariance. Subsequently, Ref.~\cite{2013PhRvD..87l2001V} adopted three
different techniques to constrain the degree of dispersion observed in the data of four Fermi/LAT GRBs.
For the subluminal case, their constraints from GRB 090510, namely $E_{{\rm QG}, 1}>7.6E_{\rm Pl}$ and
$E_{{\rm QG}, 2}>1.3\times10^{11}~\mathrm{GeV}$ for linear and quadratic leading-order LIV-induced vacuum dispersion,
further improved previous results by a factor of $\sim2$.
On 2019 January 14, the Major Atmospheric Gamma Imaging Cherenkov (MAGIC) telescopes discovered a gamma-ray signal
above 0.2 TeV from GRB 190114C, recording the most energetic photons ever detected from a GRB \cite{2019Natur.575..455M}.
The MAGIC Collaboration used this unique observation to test the dependence of the speed of light in vacuum
on its energy \cite{PhysRevLett.125.021301}. They obtained competitive lower limits on the quadratic LIV energy scale,
namely $E_{{\rm QG}, 2}>6.3\times10^{10}~\mathrm{GeV}$ ($E_{{\rm QG}, 2}>5.6\times10^{10}~\mathrm{GeV}$) for the
subluminal (superluminal) case.

Generally, Lorentz invariance has been tested with high accuracy using the spectral lags\footnote{Spectral lag
is defined as the arrival time delay between correlated photons with different energies (or between light curves
in different energy bands).} of GRBs. A key challenge in such time-of-flight tests, however, is to distinguish an
intrinsic time lag at the source from a delay induced by LIV. Possible source-intrinsic effects
caused by the unknown emission mechanism could cancel-out or enhance the LIV-induced time delay,
which would impact the reliability of the resulting constraints on LIV.

Any effects of LIV would increase with the redshift of the source (see Eq.~(\ref{eq:tLIV})), whereas source-intrinsic effects
might be independent of the redshift. Therefore, Refs.~\cite{Ellis2003,2006APh....25..402E,2008APh....29..158E} proposed
working on a statistical sample of GRBs with different redshifts to disentangle the intrinsic time lag problem.
For each GRB, Ref.~\cite{2006APh....25..402E} extracted the spectral lag in the light curves recorded in the selected observer-frame
energy bands 25--55 and 115--320 keV. To account for the unknown intrinsic time lag, they fitted the observed arrival time delays
of 35 GRBs with the inclusion of a term $b_{\rm sf}$ specified in the GRB source frame. That is, the observed spectral lags
are fitted by two components, $\Delta t_{\rm obs}=\Delta t_{\rm LIV}+b_{\rm sf}(1+z)$, reflecting the contributions from
both the LIV and source-intrinsic effects \cite{2006APh....25..402E}. Such parametrization allows one to derive a simple
linear fitting function:
\begin{equation}
\frac{\Delta t_{\rm obs}}{1+z}=a_{\rm LV}K+b_{\rm sf}\;,
\label{eq:LIV-linear}
\end{equation}
where
\begin{equation}
K=\frac{1}{(1+z)}\int_{0}^{z}\frac{(1+z')dz'}{h(z')}
\label{eq:Kfun}
\end{equation}
is a non-linear redshift function which is related to the specific cosmological model, the slope
$a_{\rm LV}=\Delta E/(H_0 E_{\rm QG})$ is connected to the energy scale of LIV, and the intercept $b_{\rm sf}$ stands for
the unknown intrinsic time lag inherited from the sources. Also, $h(z)=\sqrt{\Omega_{\rm m}(1+z)^{3}+\Omega_{\Lambda}}$
is the dimensionless expansion rate at $z$, where the standard flat $\Lambda$CDM model with parameters $\Omega_{\rm m}=0.3$
and $\Omega_{\Lambda}=1-\Omega_{\rm m}$ is adopted. Note that the linear LIV correction ($n=1$) for the subluminal case
($-$) was considered in this work. Ref.~\cite{2008APh....29..158E} compiled the rescaled time lags $\Delta t_{\rm obs}/(1+z)$
extracted from 35 light curve pairs as functions of the variable $K$ and performed a linear regression analysis of the available
data. The linear fit corresponds to $\frac{\Delta t_{\rm obs}}{1+z}=(0.0068\pm0.0067)K-(0.0065\pm0.0046)$.
With the corresponding upper limit on the slope parameter $a_{\rm LV}$, as well as the energy difference between
two observer-frame energy bands $\Delta E$, the 95\% confidence-level lower limit on the linear LIV energy scale is
$E_{\rm QG}\geq1.4\times10^{16}$ GeV \cite{2008APh....29..158E}. Going beyond the $\Lambda$CDM cosmology,
Refs. \cite{2009CQGra..26l5007B,2015ApJ...808...78P} extended this analysis to other different cosmological models,
finding the result is only weakly sensitive to the background cosmology. Some cosmology-independent
techniques were furthermore applied to this kind of analysis \cite{2018PhLB..776..284Z,2020ApJ...890..169P}.

It is worth pointing out that Ref.~\cite{2006APh....25..402E} extracted the spectral lag of each GRB in the light curves
between two fixed observer-frame energy bands. However, due to the fact that different GRBs have different redshift
measurements, these two observer-frame energy bands actually correspond to a different pair of energy bands in the GRB
source frame \cite{2012MNRAS.419..614U}, thus potentially introducing an artificial energy dependence to the extracted
spectral lag and/or systematic uncertainties to the search for LIV-induced lags.\footnote{For those cosmological sources
with various redshifts, their observer-frame quantities can be quite different than the corresponding source-frame ones.
However, note that if we focus on the observer-frame quantities of individual cosmological sources, there is no such
a problem.} This problem can be solved by choosing two appropriate energy bands fixed in the GRB source frame and
estimating the observed time lag for two projected observer-frame energy bands by the relation $E_{\rm observer}=E_{\rm source}/(1+z)$,
Ref.~\cite{2012MNRAS.419..614U} showed that the correlation between observer-frame lags and source-frame lags for the
same GRB sample has a large scatter, supporting that the source-frame lag can not be directly represented by the
observer-frame lag. Therefore, when analyzing the LIV effects in a large sample of cosmological sources with different
redshifts, it would be good to use the spectral lags extracted in the source frame. Ref.~\cite{2015MNRAS.446.1129B}
investigated the source-frame spectral lags of 56 GRBs observed by Swift/Burst Alert Telescope (BAT). This sample has
redshifts ranging from 0.35 (GRB 061021) to 5.47 (GRB 060927). For each GRB, they extracted light curves for two
observer-frame energy bands corresponding to the fixed source-frame energy bands 100--150 and 200--250 keV. These two
specific source-frame energy bands were chosen so that after transforming to the observer frame (based on the redshift
of each GRB, i.e., $[100-150]/(1+z)$ and $[200-250]/(1+z)$ keV) they still lie in the detectable energy range
of Swift/BAT. Finally, for each light-curve pairs, Ref.~\cite{2015MNRAS.446.1129B} used the improved cross-correlation
function analysis technology to estimate the spectral lag. Note that the energy difference between the mid-points of the
two source-frame energy bands is fixed at 100 keV, whereas in the observer frame (as expected), the energy difference
varies depending on the redshift of each GRB. For example, the energy difference is 16 keV in GRB 060927 and 74 keV
in GRB 061021. This is in contrast to the spectral lag extractions performed in the observer frame, where the energy
difference is treated as a constant \cite{2012MNRAS.419..614U}. Ref.~\cite{2017ApJ...851..127W} first took advantage
of the source-frame spectral lags of 56 GRBs presented in Ref.~\cite{2015MNRAS.446.1129B} to study the LIV effects.
For the subluminal case, the linear LIV-induced time delay between two observer-frame energy bands with the energy
difference $\Delta E$ is given by
\begin{equation}
\Delta t_{\rm LIV}=\frac{\Delta E}{H_{0}E_{\rm QG}}\int_{0}^{z}\frac{(1+z'){\rm d}z'}{h(z')}\\
                  =\frac{\Delta E'/(1+z)}{H_{0}E_{\rm QG}}\int_{0}^{z}\frac{(1+z'){\rm d}z'}{h(z')}\;,
\end{equation}
where $\Delta E'=100~\mathrm{keV}$ is the source-frame energy difference. Similar to the treatment of
Ref.~\cite{2006APh....25..402E}, one can formulate the intrinsic time delay problem in terms of linear regression:
\begin{equation}
\frac{\Delta t_{\rm src}}{1+z}=a'_{\rm LV}K'+b_{\rm sf}\;,
\end{equation}
where $\Delta t_{\rm src}$ is the extracted spectral lag for two fixed source-frame energy bands 100--150
and 200--250 keV,
\begin{equation}
K'=\frac{1}{(1+z)^{2}}\int_{0}^{z}\frac{(1+z')dz'}{h(z')}
\end{equation}
is a function of the redshift, and $a'_{\rm LV}=\Delta E'/(H_0 E_{\rm QG})$ is the slope in $K'$.
In order to probe the LIV effects, Ref.~\cite{2017ApJ...851..127W} performed a linear fit to the $\Delta t_{\rm src}/(1+z)$
versus $K'(z)$ data. With the optimized $a'_{\rm LV}$ and the source-frame energy difference $\Delta E'$,
the 95\% confidence-level lower limit on the linear LIV energy scale is $E_{\rm QG}\geq2.0\times10^{14}$ GeV
\cite{2017ApJ...851..127W}. This is a step forward in the investigation of LIV effects, since all previous studies
used arbitrary observer-frame energy bands.

Another point to note is that in the treatment of Ref.~\cite{2006APh....25..402E} an unknown constant was assumed
to be the intrinsic time lag in the linear fitting function (see Eq.~\ref{eq:LIV-linear}). That is, their work
based on the assumption that all GRBs have the same intrinsic time delay. However, since the durations of GRBs
span about five or six orders of magnitude, it is unlikely that high-energy photons emitted from different GRBs
(or from the same GRB) have the same intrinsic time lag relative to the emission time of low-energy photons
\cite{2016ChPhC..40d5102C}. As an improvement, Ref.~\cite{2012APh....36...47C} estimated the intrinsic time delay
between high- and low-energy photons emitted from each GRB by using the magnetic jet model. However, the magnetic
jet model relies on some given theoretical parameters, and this introduces uncertainties on the LIV results.

\begin{figure}
\centerline{\includegraphics[angle=0,width=0.6\textwidth]{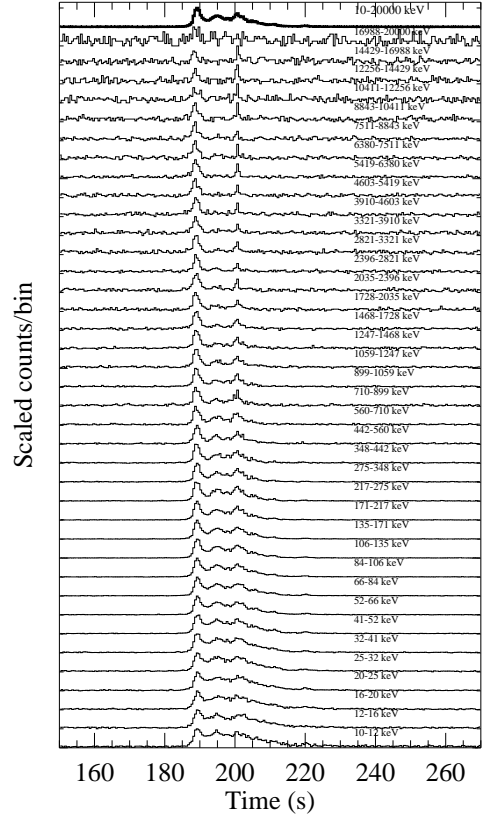}}
\vskip-0.1in
\caption{Multi-band light curves of the second sub-burst of GRB 160625B.
The top thick line represents the full-range (10--20000 keV) light curve.
Figure taken from \cite{2017ApJ...834L..13W}.}
\label{f1}
\end{figure}

\begin{figure}
\vskip-0.1in
\centerline{\includegraphics[angle=0,width=0.8\textwidth]{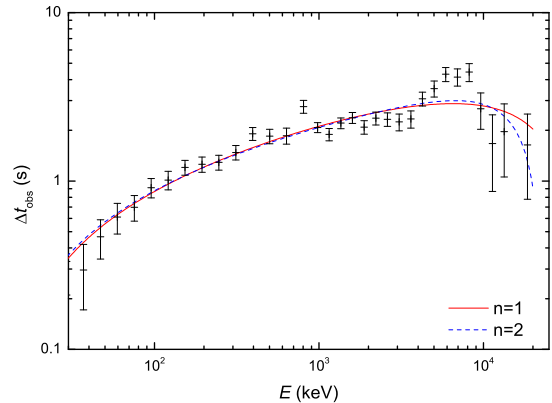}}
\vskip-0.1in
\caption{Energy dependence of the observed spectral lag $\Delta t_{\rm obs}$ of GRB 160625B relative to the softest energy band.
The best-fit theoretical curves correspond to the linear ($n=1$) and quadratic ($n=2$) LIV models, respectively.
Figure taken from \cite{2017ApJ...834L..13W}.}
\label{f2}
\end{figure}

In 2017, Ref.~\cite{2017ApJ...834L..13W} used multiple spectral lags from GRB 160625B to constrain LIV in
the photon sector. GRB 160625B was detected by the Fermi satellite on 2016 June 25, consisting of three different
isolated sub-bursts with unusually high photon statistics. Since the second sub-burst of GRB 160625B is especially bright,
it is easy to extract its high-quality light curves in different energy bands (see Figure~\ref{f1}). With multi-photon energy bands,
Ref.~\cite{2017ApJ...834L..13W} calculated the spectral lags in the light curves recorded in the lowest-energy band
(10--12 keV) relative to any other light curves with higher-energy bands, finding that the observed spectral lag
$\Delta t_{\rm obs}$ increases at $E\leq8$ MeV and then gradually decreases in the energy range $8\,{\rm MeV}\leq  E\leq20$ MeV.
The lag behavior is very peculiar, being an obvious transition from positive to negative spectral lags discovered within a
burst (see Figure~\ref{f2}). Conventionally, a positive spectral lag means an earlier arrival time for photons
with higher energies, whereas a negative spectral lag corresponds to a delayed arrival of the higher-energy photons.
Ref.~\cite{2017ApJ...834L..13W} proposed that the spectral-lag transition of GRB 160625B provides a great opportunity
to distinguish LIV-induced propagation effects from any source-intrinsic time delay in the emission of photons
at different energy bands. Since the time delay $\Delta t_{\rm LIV}$ caused by LIV may probably be accompanied by
an intrinsic energy-dependent time delay $\Delta t_{\rm int}$ arise from the unknown emission mechanism of the source,
the observed spectral lag should be comprised of two contributions,
\begin{equation}
\Delta t_{\rm obs}=\Delta t_{\rm int} + \Delta t_{\rm LIV} \;.
\end{equation}
As the observed spectral lags of most GRBs have a positive energy dependence (i.e., high-energy photons arrive earlier
than low-energy ones; see \cite{2017ApJ...844..126S,2018ApJ...865..153L}), Ref.~\cite{2017ApJ...834L..13W} suggested
that the observer-frame correlation between the intrinsic time lag and the energy $E$ can be expressed approximately by
a power-law with positive dependence,
\begin{equation}
\Delta t_{\rm int}(E)=\tau\left[\left(\frac{E}{\rm keV}\right)^{\alpha}-\left(\frac{E_{0}}{\rm keV}\right)^{\alpha}\right]\;{\rm s} \;,
\end{equation}
where $\tau>0$ and $\alpha>0$, and $E_{0}=11.34$ keV is the median value of the reference lowest-energy band (10--12 keV).
Also, when the subluminal case ($-$) is considered, the LIV effects predict a negative spectral lag (i.e.,
high-energy photons travel slower than low-energy ones). As the Lorentz-violating term takes the lead at higher energies,
the positive correlation of the lag with energy would gradually become an anti-correlation. The contributions from
both the intrinsic energy-dependent time lag and the LIV-induced time delay can therefore result in the observed lag
behavior with an apparent transition from positive to negative lags \cite{2017ApJ...834L..13W}. By directly fitting
the spectral lag behavior of GRB 160625B, Ref.~\cite{2017ApJ...834L..13W} obtained both comparatively robust limits
on the QG energy scales and a reasonable formulation of the intrinsic energy-dependent time lag. The best-fit
theoretical curves for the linear and quadratic LIV models are shown in Figure~\ref{f2}. The $1\sigma$ confidence-level
lower limits on the linear and quadratic LIV energy scales are $E_{{\rm QG},1}>0.5\times10^{16}~\mathrm{GeV}$ and
$E_{{\rm QG}, 2}>1.4\times10^{7}~\mathrm{GeV}$, respectively. The multi-photon spectral-lag data of GRB 160625B has
also been used to constrain the Lorentz-violating coefficients of the Standard Model Extension \cite{2017ApJ...842..115W}.
Recently, Ref.~\cite{2021ApJ...906....8D} showed decisive evidence that GRB 190114C is an additional burst having
a well-defined transition from positive to negative lags, providing another opportunity not only to disentangle
the intrinsic time lag problem but also to place robust constraints on LIV. Although the spectral-lag transitions of GRB 160625B
and GRB 190114C do not currently have the best sensitivity to LIV constraints, the presented method, when applied to
future bright short GRBs with similar lag features, may derive more stringent constraints on LIV.

\subsubsection{Active galactic nuclei}
Because of their rapid flux variability (ranging from minutes to years), very-high-energy (VHE, $E\geq100~\mathrm{GeV}$) emissions,
and large distances, flaring AGNs have also been considered as promising sources for LIV studies. It should be underlined that
searching for LIV-induced vacuum dispersions using both GRBs and VHE flares of AGNs is of great fundamental interest.
GRBs can be detected up to very high redshifts ($z\sim8$), but with very limited statistics of photons above a few tens of GeV.
Conversely, AGN flares can be observed by ground-based detectors with large statistics of photons up to a few tens of TeV.
Due to the absorption of VHE gamma-rays by extragalactic background light, TeV measurements are limited to sources with relatively
low redshifts. Thus, GRBs and AGNs are complementary to each other in testing LIV, and they enable us to test different energy
and redshift ranges.

There have been some competitive constraints on LIV-induced vacuum dispersions using VHE gamma-ray observations of bright AGN flares,
including the Whipple analysis of the flare of Mrk 421 \cite{1999PhRvL..83.2108B}, the H.E.S.S. analyses of the flares of
PKS 2155-304 \cite{2008PhRvL.101q0402A,2011APh....34..738H} and PG 1553+113 \cite{2015ApJ...802...65A}, and the MAGIC and H.E.S.S.
analyses of the flares of Mrk 501 \cite{2008PhLB..668..253M,2009APh....31..226M,2019ApJ...870...93A}. For the linear LIV effect
considering a subluminal case, the limit obtained from the PKS 2155-304 flare data by H.E.S.S. is the best time-of-flight limit
obtained with an AGN, yielded $E_{{\rm QG}, 1}>2.1\times10^{18}~\mathrm{GeV}$ \cite{2011APh....34..738H}. For the quadratic LIV
effect, the limits obtained from H.E.S.S.'s observation of the Mrk 501 flare are the most stringent constraints ever obtained
with an AGN, namely $E_{{\rm QG}, 2}>8.5\times10^{10}~\mathrm{GeV}$ ($E_{{\rm QG}, 2}>7.3\times10^{10}~\mathrm{GeV}$) for the subluminal
(superluminal) case \cite{2019ApJ...870...93A}.

\subsubsection{Pulsars}
Gamma-ray pulsars are a third class of astrophysical sources that are often used for LIV time-of-flight tests.
There are several reasons why they are a tempting target for these types of studies. First, due to the stable and
cyclical nature of pulsar emission, sensitivity to LIV can be systematically planned and improved by observing longer.
Second, as the timing of the pulsar is widely studied throughout the electromagnetic spectrum,
energy-dependent time delays caused by propagation effects can be more easily distinguished from source-intrinsic effects.
Moreover, although being observed several orders of magnitude closer than GRBs or AGNs, the detection of VHE emission from
pulsars will compensate for a short distance when it comes to testing the quadratic LIV term.

First limits on LIV using gamma-ray emission from the galactic Crab pulsar were obtained from the observation of
CGRO/EGRET at energies above 2 GeV \cite{Kaaret1999}, and then improved by VERITAS using VHE gamma-rays reaching up
to 120 GeV \cite{2011ICRC....7..256O,2013ICRC...33.2768Z}. Remarkably, Ref.~\cite{2017ApJS..232....9M} used
the Crab pulsar emission observed up to TeV energies by MAGIC to put new constraints on the energy scale $E_{\rm QG}$.
The 95\% confidence-level lower limits for the subluminal (superluminal) case are $E_{{\rm QG}, 1}>5.5\times10^{17}~\mathrm{GeV}$
($E_{{\rm QG}, 1}>4.5\times10^{17}~\mathrm{GeV}$) and $E_{{\rm QG}, 2}>5.9\times10^{10}~\mathrm{GeV}$
($E_{{\rm QG}, 2}>5.3\times10^{10}~\mathrm{GeV}$) for linear and quadratic LIV effects, respectively
\cite{2017ApJS..232....9M}.

\begin{table*}
\caption{A census of lower bounds on the energy scale $E_{\rm QG}$ for linear ($n=1$) and quadratic ($n=2$) LIV
for the subluminal ($-$) and superluminal ($+$) cases. These limits were produced using the time-of-flight measurements
of various astrophysical sources.}
\scriptsize
\begin{tabular}{llllll}
\hline
\hline
Source(s) & Instrument & Technique &   $E_{{\rm QG}, 1}$ [GeV]    & $E_{{\rm QG}, 2}$ [GeV]    & Refs. \\
\hline
\multicolumn{6}{c}{GRB}\\
\hline
9 GRBs$^{a}$  &  BATSE+OSSE  &  Wavelets  & ($-$) $6.9\times10^{15}$  & ($-$) $2.9\times10^{6}$  &\cite{Ellis2003}  \\
GRB 021206$^{b}$  &  RHESSI  &  Peak times at different energies & ($-$) $1.8\times10^{17}$ &  & \cite{2004ApJ...611L..77B}  \\
15 GRBs  &  HETE-2  &  Wavelets  & ($-$) $2.0\times10^{15}$ &  & \cite{2008ApJ...676..532B}  \\
11 GRBs  &  INTEGRAL  &  Likelihood  & ($-$) $3.2\times10^{11}$ &  & \cite{2008GReGr..40.1731L}   \\
GRB 051221A  &  Konus-Wind+Swift  &  Peak times at different energies  & ($-$) $6.6\times10^{16}$ & ($-$) $6.2\times10^{6}$ & \cite{2006JCAP...05..017R}   \\
\multirow{2}{*}{35 GRBs$^{c}$}  &  BATSE+HETE-2  &  \multirow{2}{*}{Observer-frame spectral lags}  & \multirow{2}{*}{($-$) $1.4\times10^{16}$} &  & \multirow{2}{*}{\cite{2006APh....25..402E,2008APh....29..158E}}  \\
      & +Swift  &   &   &  &   \\
56 GRBs  &  Swift &  Source-frame spectral lags & ($-$) $2.0\times10^{14}$ &   & \cite{2017ApJ...851..127W}  \\
\multirow{2}{*}{GRB 080916C}  &  \multirow{2}{*}{Fermi GBM+LAT}  &  Associating a 13.2 GeV photon with   & \multirow{2}{*}{($-$) $1.3\times10^{18}$} &  & \multirow{2}{*}{\cite{2009Sci...323.1688A}}   \\
      &   &  the trigger time &   &  &  \\
\multirow{2}{*}{GRB 090510}  &  \multirow{2}{*}{Fermi GBM+LAT}  &  Associating a 31 GeV photon with   & \multirow{2}{*}{($-$) $1.5\times10^{19}$} &  & \multirow{2}{*}{\cite{2009Natur.462..331A}}   \\
      &   &  the start of the first GBM pulse &   &  &  \\
\multirow{2}{*}{GRB 090510}   &  \multirow{2}{*}{Fermi/LAT}  &  PairView+Likelihood  & ($-$) $9.3\times10^{19}$ & ($-$) $1.3\times10^{11}$  & \multirow{2}{*}{\cite{2013PhRvD..87l2001V}}   \\
      &   &  +Sharpness-Maximization Method & ($+$) $1.3\times10^{20}$  & ($+$) $9.4\times10^{10}$ & \\
\multirow{2}{*}{GRB 190114C}  &  \multirow{2}{*}{MAGIC}  &  \multirow{2}{*}{Likelihood}  & ($-$) $5.8\times10^{18}$  & ($-$) $6.3\times10^{10}$ & \multirow{2}{*}{\cite{PhysRevLett.125.021301}}   \\
      &   &   & ($+$) $5.5\times10^{18}$  & ($+$) $5.6\times10^{10}$ & \\
GRB 160625B  &  Fermi/GBM  &  Spectral-lag transition & ($-$) $0.5\times10^{16}$ & ($-$) $1.4\times10^{7}$ & \cite{2017ApJ...834L..13W}   \\
GRB 190114C  &  Fermi/GBM  &  Spectral-lag transition & ($-$) $2.2\times10^{14}$ & ($-$) $0.9\times10^{6}$ & \cite{2021ApJ...906....8D}   \\
\hline
\multicolumn{6}{c}{AGN}\\
\hline
Mrk 421  &  Whipple  &  Binning  & ($-$) $4.0\times10^{16}$ &   & \cite{1999PhRvL..83.2108B}   \\
PKS 2155-304 &  H.E.S.S. &  Modified cross correlation function  & ($-$) $7.2\times10^{17}$ &  ($-$) $1.4\times10^{9}$  & \cite{2008PhRvL.101q0402A}   \\
PKS 2155-304 &  H.E.S.S. &  Likelihood  & ($-$) $2.1\times10^{18}$ & ($-$) $6.4\times10^{10}$ & \cite{2011APh....34..738H}  \\
\multirow{2}{*}{PG 1553+113}  &  \multirow{2}{*}{H.E.S.S.} &  \multirow{2}{*}{Likelihood}  & ($-$) $4.1\times10^{17}$ & ($-$) $2.1\times10^{10}$ & \multirow{2}{*}{\cite{2015ApJ...802...65A}}  \\
      &   &   & ($+$) $2.8\times10^{17}$  & ($+$) $1.7\times10^{10}$ & \\
Mrk 501  &  MAGIC &  Energy cost function  & ($-$) $2.1\times10^{17}$ & ($-$) $2.6\times10^{10}$ & \cite{2008PhLB..668..253M}   \\
Mrk 501  &  MAGIC &  Likelihood  & ($-$) $3.0\times10^{17}$  & ($-$) $5.7\times10^{10}$ & \cite{2009APh....31..226M}  \\
\multirow{2}{*}{Mrk 501}  &  \multirow{2}{*}{H.E.S.S.} &  \multirow{2}{*}{Likelihood}  & ($-$) $3.6\times10^{17}$  & ($-$) $8.5\times10^{10}$  & \multirow{2}{*}{\cite{2019ApJ...870...93A}}   \\
     &   &   & ($+$) $2.6\times10^{17}$  & ($+$) $7.3\times10^{10}$ & \\
\hline
\multicolumn{6}{c}{Pulsar}\\
\hline
\multirow{2}{*}{Crab}  &  \multirow{2}{*}{CGRO/EGRET}  &  Pulse arrival times   & \multirow{2}{*}{($-$) $1.8\times10^{15}$} &   & \multirow{2}{*}{\cite{Kaaret1999}}   \\
      &   & in different energy bands  &  &  &  \\
Crab    &  VERITAS  &  Likelihood    & ($-$) $3.0\times10^{17}$  & ($-$) $7.0\times10^{9}$   & \cite{2011ICRC....7..256O}   \\
\multirow{2}{*}{Crab}    &  \multirow{2}{*}{VERITAS }  &  \multirow{2}{*}{Dispersion Cancellation}    & ($-$) $1.9\times10^{17}$ &  & \multirow{2}{*}{\cite{2013ICRC...33.2768Z}}   \\
     &   &   & ($+$) $1.7\times10^{17}$  &  & \\
\multirow{2}{*}{Crab}    &  \multirow{2}{*}{MAGIC}  &  \multirow{2}{*}{Likelihood}    & ($-$) $5.5\times10^{17}$  & ($-$) $5.9\times10^{10}$ & \multirow{2}{*}{\cite{2017ApJS..232....9M}}   \\
     &   &   & ($+$) $4.5\times10^{17}$  & ($+$) $5.3\times10^{10}$ & \\
\hline
\end{tabular}
\label{table1}
\\
$^{a}$Limit obtained taking no account of the factor $(1+z)$ in the integral of Eq.~(\ref{eq:Kfun}).
\\
$^{b}$The spectral and temporal properties of GRB 021206 were used to estimate the pseudo redshift.
\\
$^{c}$The Limits of Ref.~\cite{2006APh....25..402E} were corrected in Ref.~\cite{2008APh....29..158E}
by taking account of the factor $(1+z)$ in the integral of Eq.~(\ref{eq:Kfun}).
\end{table*}

The most important results obtained to date with vacuum dispersion time-of-flight measurements of various astrophysical
events are listed in Table~\ref{table1}. The most stringent limits so far on $E_{\rm QG}$ have been obtained with
Fermi/LAT's observation of GRB 090510. The limits set are $E_{{\rm QG}, 1}>9.3\times10^{19}~\mathrm{GeV}$
($E_{{\rm QG}, 1}>1.3\times10^{20}~\mathrm{GeV}$) for a linear, and $E_{{\rm QG}, 2}>1.3\times10^{11}~\mathrm{GeV}$
($E_{{\rm QG}, 2}>9.4\times10^{10}~\mathrm{GeV}$) for a quadratic LIV, for the subluminal (superluminal) case
\cite{2013PhRvD..87l2001V}. Obviously, these time-of-flight studies using gamma-ray emissions in the GeV--TeV range
provide us at present with the best opportunity to search for Planck-scale modifications of the photon dispersion
relation. Nevertheless, while they offer meaningful constraints for the linear ($n=1$) modification,
they are still much weaker for deviations that appear at the quadratic ($n=2$) order.

\section{Vacuum birefringence}
\subsection{General formulae}
The Charge-Parity-Time (CPT) theorem states that the physical laws are invariant under charge conjugation, parity transformation,
and time reversal. In some QG theories that invoke LIV, the CPT theorem no longer holds. In the absence of Lorentz symmetry,
the CPT invariance, if needed, should be imposed as an additional assumption of the theory. In the photon sector, these theories
invoke a Lorentz- and CPT-violating dispersion relation of the form \cite{2003PhRvL..90u1601M}
\begin{equation}\label{eq:dispersion}
  E_{\pm}^2=p^2c^2\pm \frac{2\eta}{E_{\rm pl}} p^3c^3\;,
\end{equation}
where the sign $\pm$ corresponds to the helicity, i.e., two different circular polarization states, and $\eta$ is a dimensionless
parameter that needs to be bound. Note that the parameter $\eta$ exactly vanishes in LIV but CPT invariant theories.
Therefore, in this sense, such tests might be less broad and general than those tests based on vacuum dispersion.

The linear polarization can be decomposed into right- or left-handed circular polarization states. If $\eta\neq0$,
then the dispersion relation (\ref{eq:dispersion}) indicates that photons with opposite circular polarizations have
slightly different phase velocities and therefore propagate with different speeds, leading to an energy-dependent rotation of
the polarization vector of a linearly polarized light. This effect is known as vacuum birefringence.
The rotation angle during the propagation from the source at redshift $z$ to the observer
is expressed as \cite{2011PhRvD..83l1301L,2012PhRvL.109x1104T}
\begin{equation}\label{eq:theta-LIV}
  \Delta\phi_{\rm LIV}(E)\simeq\eta\frac{E^2 F(z)}{\hbar E_{\rm pl}H_{0}}\;,
\end{equation}
where $E$ is the observed photon energy, and $F(z)$ is a function related to the cosmological model, which reads
(in the standard flat $\Lambda$CDM model)
\begin{equation}
F(z)=\int_0^z\frac{\left(1+z'\right)dz'}{\sqrt{\Omega_{\rm m}\left(1+z'\right)^3+\Omega_{\Lambda}}}\;.
\end{equation}

Generally, it is hard to know the intrinsic polarization angles for photons emitted with different energies from
a given astrophysical source. If one possessed this information, evidence for vacuum birefringence (i.e.,
an energy-dependent rotation of the polarization plane of linearly polarized photons) could be directly obtained by
measuring differences between the known intrinsic polarization angles and the actual observed polarization angles
at different energies. Even in the absence of such knowledge, however, the birefringent effect can still be
tested by polarized sources at arbitrary redshifts. This is because a large degree of birefringence would add
opposite oriented polarization vectors, effectively washing out most, if not all, of the net polarization of
the signal. Therefore, the polarization detections with high significance levels can put upper bounds on
the energy-dependent birefringent effect.

\subsection{Present constraints from polarization measurements}
Observations of linear polarization from distant astrophysical sources can be used to place upper bounds on the
birefringent parameter $\eta$. The vacuum birefringence constraints stem from the fact that if the rotation angles
of photons with different energies (Eq.~\ref{eq:theta-LIV}) differ by more than $\pi/2$ over an energy range ($E_1<E<E_2$),
then the net polarization of the signal would be significantly depleted and could not be as high as the observed level.
Thus, the detection of highly polarized photons implies that the differential rotation angle $|\Delta\phi(E_{2})-\Delta\phi(E_{1})|$
should be smaller than $\pi/2$, i.e.,
\begin{equation}
|\Delta\phi(E_{2})-\Delta\phi(E_{1})|=\eta\frac{\left(E_{2}^2-E_{1}^2\right) F(z)}{\hbar E_{\rm pl}H_{0}}\leq\frac{\pi}{2}\;.
\end{equation}

Previously, Ref.~\cite{2001PhRvD..64h3007G} used the observed linear polarization of ultraviolet light from
a galaxy 3C 256 at a distance of around 300 Mpc to set an upper bound of $\eta<10^{-4}$. Ref.~\cite{2008PhRvD..78j3003M}
derived a tighter constraint of $\eta<9\times10^{-10}$ using the hard X-ray polarization observation from the Crab Nebula.
Much stronger limits of $\eta<10^{-14}$ were obtained by Refs.~\cite{2003Natur.426Q.139M,2004PhRvL..93b1101J}
using a report of polarized gamma-rays observed in the prompt emission from GRB 021206 \cite{2003Natur.423..415C}.
However, this claimed polarization detection has been refuted by subsequent re-analyses of the same data
\cite{2004MNRAS.350.1288R,2004ApJ...613.1088W}. Using the reported detection of polarized soft gamma-ray emission
from GRB 041219A \cite{2007ApJS..169...75K,2007A&A...466..895M,2009ApJ...695L.208G},
Refs.~\cite{2011PhRvD..83l1301L,2011APh....35...95S} obtained a stringent upper limit on $\eta$ of $1.1\times10^{-14}$
and $2.4\times10^{-15}$, respectively. But again, the previous reports of the gamma-ray polarimetry for GRB 041219A
have been disputed (see \cite{2012PhRvL.109x1104T} for more explanations), and the arguments for the resulting constraints
given by Refs.~\cite{2011PhRvD..83l1301L,2011APh....35...95S} are still open to questions.

Contrary to those controversial reports, the detection of gamma-ray linear polarization by the Gamma-ray burst
Polarimeter (GAP) onboard the Interplanetary Kite-craft Accelerated by Radiation Of the Sun (IKAROS) is fairly credible
and thus can be used to set a reliable limit on the birefringent parameter \cite{2012PhRvL.109x1104T}. IKAROS/GAP
clearly detected linear polarizations in the prompt gamma-ray emission of three GRBs, with a polarization degree
of $\Pi=27\pm11\%$ for GRB 100826A \cite{2011ApJ...743L..30Y}, $\Pi=70\pm22\%$ for GRB 110301A \cite{2012ApJ...758L...1Y},
and $\Pi=84^{+16}_{-28}\%$ for GRB 110721A \cite{2012ApJ...758L...1Y}. The detection significance is $2.9\sigma$,
$3.7\sigma$, and $3.3\sigma$, respectively. With the assumption that $|\Delta\phi(E_{2})-\Delta\phi(E_{1})|\leq \pi/2$,
Ref.~\cite{2012PhRvL.109x1104T} applied the polarization data of these three GRBs to derive a severe upper bound on
$\eta$ in the order of $\mathcal{O}(10^{-15})$. Note that since the redshifts of these GRBs were not measured,
Ref.~\cite{2012PhRvL.109x1104T} used a luminosity indicator for GRBs to estimate the possible redshifts.
Utilizing the real redshift determination ($z=1.33$) together with the polarimetric data of GRB 061122,
Ref.~\cite{2013MNRAS.431.3550G} derived a more stringent limit ($\eta<3.4\times10^{-16}$) on the possibility of LIV
based on the vacuum birefringent effect. The current deepest limit of $\eta<1.0\times10^{-16}$ was obtained
by Ref.~\cite{2014MNRAS.444.2776G} using the most distant polarized burst GRB 140206A at redshift $z=2.739$.

All the polarization constraints presented above were based on the assumption that the observed polarization degree
would be severely suppressed for a given energy range if the differential rotation angle $|\Delta\phi(E_{2})-\Delta\phi(E_{1})|$
is too large, regardless of the intrinsic polarization fraction at the corresponding rest-frame energy range.
However, Ref.~\cite{2016MNRAS.463..375L} gave a detailed calculation on the evolution of GRB polarization
induced by the vacuum birefringent effect, and showed that, even if $|\Delta\phi(E_{2})-\Delta\phi(E_{1})|$
is as large as $\pi/2$, more than 60\% of the initial polarization degree can be conserved. This is in conflict with
the general perception that $|\Delta\phi(E_{2})-\Delta\phi(E_{1})|$ could not be larger than $\pi/2$ when high polarization
is observed. Thus, Ref.~\cite{2016MNRAS.463..375L} suggested that it is inappropriate to simply use $\pi/2$ as
the upper limit of $|\Delta\phi(E_{2})-\Delta\phi(E_{1})|$ to constrain the vacuum birefringent effect.
Applying their formulae for calculating the polarization evolution to two true GRB events, Ref.~\cite{2016MNRAS.463..375L}
obtained the most conservative limits on the birefringent parameter from the polarimetric data of GRB 061122
and GRB 110721A as $\eta<0.5\times10^{-16}$ and $\eta<4.0\times10^{-16}$, respectively. Following the analysis method
proposed in Ref.~\cite{2016MNRAS.463..375L} and using the latest detections of prompt emission polarization in GRBs,
Ref.~\cite{2019MNRAS.485.2401W} improved existing bounds on a deviation from Lorentz invariance through the vacuum
birefringent effect by factors ranging from two to ten.

\begin{figure}
\centering
\vskip-0.2in
\includegraphics[width=1.0\textwidth]{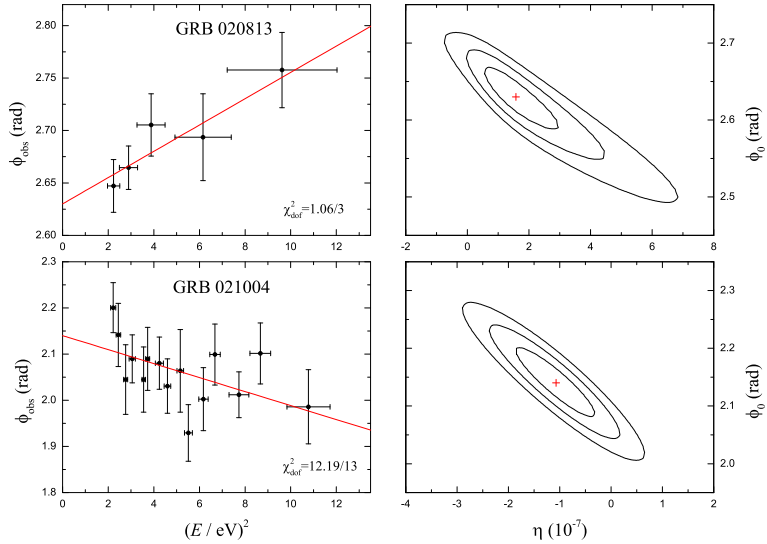}
\vskip-0.2in
\caption{Fit to multiband polarization observations of the optical afterglows of GRB 020813 and GRB 021004.
Left panels: linear fits of the spectropolarimetric data.
Right panels: 1-3$\sigma$ confidence levels in the $\phi_{0}$-$\eta$ plane. Reproduced from \cite{2020EPJP..135..527W}.}
\label{f3}
\end{figure}

Instead of requiring the more complicated and indirect assumption that $|\Delta\phi(E_{2})-\Delta\phi(E_{1})|\leq\pi/2$,
some authors simply assume that all photons in the observed energy band are emitted with the same (unknown)
intrinsic polarization angle \cite{2007MNRAS.376.1857F,2020EPJP..135..527W,2021Galax...9...44Z}.
If the rotation angle of the linear polarization plane arising from the birefringent effect
$\Delta\phi_{\rm LIV}(E)$ is considered here, the observed linear polarization angle at a certain $E$ with
an intrinsic polarization angle $\phi_{0}$ should be
\begin{equation}\label{eq:phi_obs}
\phi_{\rm obs}=\phi_{0}+\Delta\phi_{\rm LIV}\left(E\right)\;.
\end{equation}
Since $\phi_{0}$ is assumed to be an unknown constant, we expect to observe the birefringent effect as an energy-dependent
linear polarization vector. Such an observation could give a robust limit on the birefringent parameter $\eta$.
Ref.~\cite{2007MNRAS.376.1857F} searched for the energy-dependent change of the polarization angle in the
spectropolarimetric observations of the optical afterglows of GRB 020813 and GRB 021004. By fitting the multiband
polarimetric data of these two GRBs with Eq.~(\ref{eq:phi_obs}), Ref.~\cite{2007MNRAS.376.1857F}
obtained constraints on both $\phi_{0}$ and $\eta$ (see Figure~\ref{f3}). At the $3\sigma$ confidence level,
the joint constraint on $\eta$ from two GRBs is $-2\times10^{-7}<\eta<1.4\times10^{-7}$ (see also \cite{2020EPJP..135..527W}).
Ref.~\cite{2021Galax...9...44Z} applied the same analysis method to multiband optical polarimetry of five blazars
and obtained $\eta$ constraints with similar accuracy. It is clear from Eq.~(\ref{eq:theta-LIV}) that the larger the distance of
the polarized source, and the higher the energy band of the polarization observation, the greater the sensitivity to
small values of $\eta$. As expected, less stringent constraints on $\eta$ were obtained from the optical polarization data
\cite{2007MNRAS.376.1857F,2020EPJP..135..527W,2021Galax...9...44Z}.

\begin{table*}
\caption{A census of limits on the vacuum birefringent parameter $\eta$.
These limits were obtained from the linear polarization measurements of various astrophysical sources.}
\scriptsize
\begin{tabular}{lllcccc}
\hline
\hline
Source & Distance & Polarimeter & Energy Rang$^{a}$ &  $\Pi$ & $\eta$    & Refs. \\
\hline
Galaxy 3C 256    & $d=300$ Mpc & Spectropolarimeter & Ultraviolet & $16.4\pm2.2\%$ & $<10^{-4}$ & \cite{2001PhRvD..64h3007G} \\
Crab Nebula      & $d=1.9$ kpc  & INTEGRAL/SPI & 100--1000 keV & $46\pm10\%$ & $<9\times10^{-10}$ & \cite{2008PhRvD..78j3003M} \\
\hline
\multicolumn{7}{c}{GRB}\\
\hline
GRB 021206       & $d=10^{10}$ light years$^{c}$ & RHESSI & 150--2000 keV & $80\pm20\%$$^{b}$ & $<10^{-14}$ & \cite{2003Natur.426Q.139M} \\
GRB 021206       & $d=0.5$ Gpc$^{d}$ &RHESSI & 150--2000 keV & $80\pm20\%$$^{b}$ & $<0.5\times10^{-14}$ & \cite{2004PhRvL..93b1101J} \\
GRB 041219A      & $z=0.02$$^{e}$  & INTEGRAL/IBIS & 200--800 keV & $43\pm25\%$$^{b}$ & $<1.1\times10^{-14}$ & \cite{2011PhRvD..83l1301L} \\
GRB 041219A      & $z=0.2$$^{f}$  & INTEGRAL/SPI & 100--350 keV & $96\pm40\%$$^{b}$ & $<2.4\times10^{-15}$ & \cite{2011APh....35...95S}\\
GRB 100826A      & $z=0.71$$^{f}$  & IKAROS/GAP & 70--300 keV & $27\pm11\%$ & $<2.0\times10^{-14}$ & \cite{2012PhRvL.109x1104T}\\
GRB 110301A      & $z=0.21$$^{f}$ & IKAROS/GAP & 70--300 keV & $70\pm22\%$ & $<1.0\times10^{-14}$ &  \cite{2012PhRvL.109x1104T}\\
GRB 110721A      & $z=0.45$$^{f}$ & IKAROS/GAP & 70--300 keV & $84^{+16}_{-28}\%$ & $<2.0\times10^{-15}$ & \cite{2012PhRvL.109x1104T}\\
GRB 061122       & $z=1.33$   & INTEGRAL/IBIS & 250--800 keV & $>60\%$ & $<3.4\times10^{-16}$ & \cite{2013MNRAS.431.3550G}\\
GRB 140206A      & $z=2.739$  & INTEGRAL/IBIS & 200--400 keV & $>48\%$ & $<1.0\times10^{-16}$ & \cite{2014MNRAS.444.2776G}\\
GRB 061122       & $z=1.33$   & INTEGRAL/IBIS & 250--800 keV & $>60\%$ & $<0.5\times10^{-16}$ & \cite{2016MNRAS.463..375L}\\
GRB 110721A      & $z=0.45$$^{f}$  & IKAROS/GAP & 70--300 keV & $84^{+16}_{-28}\%$ & $<4.0\times10^{-16}$ & \cite{2016MNRAS.463..375L}\\
GRB 061122	     & $z=1.33$   & INTEGRAL/IBIS	&	250--800 keV	&$	>60\%	$	&	$<0.5\times10^{-16}$ & \cite{2019MNRAS.485.2401W}\\
GRB 100826A      & $z=0.083$$^{f}$	& IKAROS/GAP	&	70--300 keV	&$	27\pm11\%	$ &	 $1.2^{+1.4}_{-0.7}\times10^{-14}$ & \cite{2019MNRAS.485.2401W}\\
GRB 110301A      & $z=0.082$$^{f}$	& IKAROS/GAP	&	70--300	keV &$	70\pm22\%	$&	 $4.3^{+5.4}_{-2.3}\times10^{-15}$ & \cite{2019MNRAS.485.2401W}\\
GRB 110721A	     & $z=0.382$   & IKAROS/GAP	&	70--300 keV &$	84^{+16}_{-28}\%$&	 $5.1^{+4.0}_{-5.1}\times10^{-16}$ & \cite{2019MNRAS.485.2401W}\\
GRB 140206A	     & $z=2.739$    & INTEGRAL/IBIS	&	200--400 keV &$	>48\%	$&	$<1.0\times10^{-16}$ & \cite{2019MNRAS.485.2401W}\\
GRB 160106A      & $z=0.091$$^{f}$	& AstroSat/CZTI	&	100--300 keV	&$	68.5	\pm	24\%$&	 $3.4^{+1.4}_{-1.8}\times10^{-15}$ & \cite{2019MNRAS.485.2401W}\\
GRB 160131A	     & $z=0.972$  & AstroSat/CZTI	&	100--300 keV	&$	94	\pm	31\%	$&	 $1.2^{+2.0}_{-1.2}\times10^{-16}$ & \cite{2019MNRAS.485.2401W}\\
GRB 160325A      & $z=0.161$$^{f}$	& AstroSat/CZTI	&	100--300 keV	&$	58.75\pm23.5\%	$&	 $2.3^{+1.0}_{-0.9}\times10^{-15}$ & \cite{2019MNRAS.485.2401W}\\
GRB 160509A	     & $z=1.17$   & AstroSat/CZTI	&	100--300 keV	&$	96	\pm	40\%	$&	 $0.8^{+2.2}_{-0.8}\times10^{-16}$ & \cite{2019MNRAS.485.2401W}\\
GRB 160802A      & $z=0.105$$^{f}$	& AstroSat/CZTI	&	100--300 keV	&$	85	\pm	29\%	$&	 $2.0^{+1.7}_{-2.0}\times10^{-15}$ & \cite{2019MNRAS.485.2401W}\\
GRB 160821A      & $z=0.047$$^{f}$	& AstroSat/CZTI	&	100--300 keV	&$	48.7\pm	14.6\%	$&	 $8.9^{+1.7}_{-1.7}\times10^{-15}$ & \cite{2019MNRAS.485.2401W}\\
GRB 160910A      & $z=0.272$$^{f}$	& AstroSat/CZTI	&	100--300 keV	&$	93.7\pm	30.92\%	$&	 $4.7^{+7.6}_{-4.7}\times10^{-16}$ & \cite{2019MNRAS.485.2401W}\\
GRB 020813       & $z=1.255$        & LRISp & 3500--8800 {\AA} & 1.8\%--2.4\% & ($-2.0$--1.4)$\times10^{-7}$ & \cite{2007MNRAS.376.1857F} \\
GRB 021004       & $z=2.328$     & VLT & 3500--8600 {\AA} & $1.88\pm0.05\%$ & ($-2.0$--1.4)$\times10^{-7}$ & \cite{2007MNRAS.376.1857F}\\
GRB 020813       & $z=1.255$        & LRISp & 3500--8800 {\AA} & 1.8\%--2.4\% & ($-1.8$--1.1)$\times10^{-7}$ & \cite{2020EPJP..135..527W} \\
GRB 021004       & $z=2.328$     & VLT & 3500--8600 {\AA} & $1.88\pm0.05\%$ & ($-1.8$--1.1)$\times10^{-7}$ & \cite{2020EPJP..135..527W}\\
\hline
\multicolumn{7}{c}{Blazar}\\
\hline
 3C 66A       & $z=0.444$     & Photopolarimeter   & $UBVRI$  & 14.1\%--33.7\% & $(3.4\pm1.2)\times10^{-7}$ & \cite{2021Galax...9...44Z}\\
S5 0716+714      & $z=0.310$     & Photopolarimeter  & $UBVRI$  & 15.1\%--20.1\% & $(-1.0\pm4.6)\times10^{-7}$ & \cite{2021Galax...9...44Z}\\
OJ 287       & $z=0.306$     & Photopolarimeter   & $UBVRI$  & 5.2\%--22.8\% & $(6.2\pm7.2)\times10^{-7}$ & \cite{2021Galax...9...44Z}\\
MK 421       & $z=0.031$     & Photopolarimeter   & $UBVRI$  & 1.7\%--3.5\% & $(6.5\pm22.8)\times10^{-6}$ & \cite{2021Galax...9...44Z}\\
PKS 2155-304       & $z=0.116$     & Photopolarimeter  &  $UBVRI$  & 3\%--7\% & $(-5.3\pm3.6)\times10^{-7}$ & \cite{2021Galax...9...44Z}\\
\hline
\end{tabular}
\label{table2}
\\
$^{a}$The energy range in which polarization is observed.
\\
$^{b}$The claimed polarization detections have been refuted.
\\
$^{c}$The distance of GRB 021206 was assumed to be $10^{10}$ light years.
\\
$^{d}$The distance of GRB 021206 was assumed to be 0.5 Gpc.
\\
$^{e}$The lower limit to the photometric redshift of GRB 041219A was adopted.
\\
$^{f}$The redshifts of these GRBs were estimated by the empirical luminosity relation.
\end{table*}

Table~\ref{table2} presents a summary of astrophysical polarization constraints on a possible LIV through
the vacuum birefringent effect. The current best limits on the birefringent parameter,
$\eta<\mathcal{O}(10^{-16})$, have been obtained by the detections of gamma-ray linear polarization of GRBs
\cite{2013MNRAS.431.3550G,2014MNRAS.444.2776G,2016MNRAS.463..375L,2019MNRAS.485.2401W}.

\section{Photon decay and photon splitting}
There are various forms of modified dispersion relation for different particles and underlying QG theories.
One of the Lorentz-violating dispersion relations for photons takes the generalized form \cite{1998Natur.393..763A}
\begin{equation}\label{eq:decay-dispersion}
  E^2-p^2=\pm|\alpha_{n}|p^{n+2} \;,
\end{equation}
where $n$ is the leading order of the modification from the underlying QG theory, $\alpha_{n}$ is the $n$th order LIV
parameter, and the sign $\pm$ refers to the superluminal ($+$) and subluminal ($-$) cases. For $n>0$, limits on
$\alpha_{n}$ can be interpreted in terms of the QG energy scale,
\begin{equation}\label{eq:E-alpha}
  E_{{\rm QG}, n} = \alpha_{n}^{-1/n}\;.
\end{equation}

\subsection{Photon decay}
In the superluminal LIV case, photons can decay into electron-positron pairs, $\gamma\rightarrow e^{+}e^{-}$.
The resulting decay rates $\Gamma_{\gamma\rightarrow e^{+}e^{-}}$ are rapid and effective at energies where
the process is allowed. Once above the energy threshold, this photon decay process would lead to a hard cutoff
in the gamma-ray spectrum without any high-energy photons being observed \cite{2017PhRvD..95f3001M}.
The threshold for any order $n$ is given by
\begin{equation}\label{eq:alpha}
\alpha_{n}\leq \frac{4m_{e}^{2}}{E^{n}\left(E^{2}-4m_{e}^{2}\right)}\;,
\end{equation}
where $m_e$ represents the electron mass \cite{2017PhRvD..95f3001M}. One can see from Eqs.~(\ref{eq:E-alpha})
and (\ref{eq:alpha}) that the upper limits on $\alpha_{n}$ (lower limits on $E_{{\rm QG}, n}$) become tighter
with the increase in the observed photon energy by a factor of $E^{-(n+2)}$ ($E^{1+2/n}$ for lower limits on
$E_{{\rm QG}, n}$).

Following Eqs.~(\ref{eq:E-alpha}) and (\ref{eq:alpha}), one then has $E_{{\rm QG}, n}$ for $n=1$ and $2$,
\begin{equation}\label{eq:decay-limit1}
E_{{\rm QG}, 1}\geq9.57\times10^{23} \;{\rm eV} \left(\frac{E}{\rm TeV}\right)^{3}\;,
\end{equation}
\begin{equation}\label{eq:decay-limit2}
E_{{\rm QG}, 2}\geq9.78\times10^{17} \;{\rm eV} \left(\frac{E}{\rm TeV}\right)^{2}\;.
\end{equation}
Therefore, a lower limit for $E_{{\rm QG}, n}$ in the photon sector can be obtained directly from any observed
high-energy photon event.

\subsection{Photon splitting}
A second superluminal LIV decay process is photon splitting to multiple photons, $\gamma\rightarrow N\gamma$.
The dominant splitting process with quadratic modification ($n=2$) is the photon decay into three photons
($\gamma\rightarrow 3\gamma$) \cite{2019JCAP...04..054A}. The decay rate of photon splitting
(in units of energy) is \cite{2019JCAP...04..054A,2019EPJC...79.1011S}
\begin{equation}
\Gamma_{\gamma\rightarrow 3\gamma}=5\times10^{-14}\frac{E^{19}}{m_{e}^{8}E_{{\rm QG}, 2}^{10}}\;,
\end{equation}
which is much smaller than the photon decay rate $\Gamma_{\gamma\rightarrow e^{+}e^{-}}$. Despite the lack of
an energy threshold, this process is kinematically allowed whenever $E^{2}>p^{2}$ (i.e., superluminal LIV).
It becomes significant when photons travel through astrophysical distances and also predicts a hard cutoff
at high photon energies in astrophysical spectra.

If we find the evidence of a cutoff in the spectrum of an energetic astrophysical source at distance $d$,
we can equate the mean free path of a photon to the source distance. That is, we take
$d \times \Gamma_{\gamma\rightarrow 3\gamma}=1$, with $\Gamma_{\gamma\rightarrow 3\gamma}$ translated
to units of $\rm kpc^{-1}$. The corresponding LIV limit is therefore given by
\begin{equation}\label{eq:splitting-limit}
E_{{\rm QG}, 2}\geq3.33\times10^{19} \;{\rm eV} \left(\frac{d}{\rm kpc}\right)^{0.1} \left(\frac{E}{\rm TeV}\right)^{1.9}\;,
\end{equation}
which is a function of the highest photon energy. Once again, this photon splitting in flight from the source
provides a direct way to constrain the quadratic LIV energy scale that mainly relies on the highest energy
photons observed.

\subsection{Present constraints from spectral cutoff}
If there is no sign of clear cutoff at the highest energy end of the photon spectra of astrophysical sources,
then stringent lower limits on the LIV energy scale can be derived. There have been some resulting constraints
on LIV-induced spectral cutoff using sub-PeV measurements of gamma-ray spectra from astrophysical sources, including
the HEGRA observations of the Crab Nebula spectrum \cite{2017PhRvD..95f3001M,2019JCAP...04..054A}, the Tibet
observations of the Crab Nebula spectrum \cite{2019EPJC...79.1011S}, the HAWC observations of the spectra of
the Crab and three other sources \cite{2020PhRvL.124m1101A}, and the LHAASO observations of the spectra of
the Crab and four other sources \cite{2021arXiv210507927C,2021arXiv210612350T}.

For instance, the LHAASO Collaboration studied the superluminal LIV effect using the observations of unprecedentedly
high-energy gamma rays, with a rigorous statistical technique and a careful assessment of the systematic uncertainties
\cite{2021arXiv210612350T}. Recently, 12 ultra-high-energy gamma-ray Galactic sources above 100 TeV have been detected
by LHAASO \cite{2021Natur.594...33C}. The highest energy photon-like event from the source LHAASO J2032+4102 is about
1.4 PeV. The second highest energy photon-like event from LHAASO J0534+2202 (Crab Nebula) is about 0.88 PeV. Since
higher energy photons may provide better limits on the LIV energy scale, the LHAASO Collaboration adopted the two
sources J2032+4102 and J0534+2202 for their purpose \cite{2021arXiv210612350T}. Two general spectral forms are assumed
for both sources, i.e., a log-parabolic form and a power-law form. Both the photon decay ($\gamma\rightarrow e^{+}e^{-}$)
and photon splitting ($\gamma\rightarrow 3\gamma$) processes predict that the energy spectrum of a source has a quasi-sharp
cutoff at the highest energy part. If LIV does exist, the expected spectrum of the source can then be expressed as
\cite{2021arXiv210612350T}
\begin{equation}
f(E)=A_{0}\left(\frac{E}{E_0}\right)^{-\alpha-\beta\ln\left(E/E_{0}\right)}H(E-E_{\rm cut})\;,
\end{equation}
where $A_{0}$ is the flux normalization, $\alpha$ and $\beta$ are the spectral indices, $E_{0}=20$ TeV is a reference energy,
$H(E-E_{\rm cut})$ stands for the Heaviside step function, and $E_{\rm cut}$ is the cutoff energy. The above expression
is for the log-parabolic spectrum. When $\beta\equiv0$, the log-parabolic spectrum reduces to the power-law spectrum.
The spectral parameters can be optimized by maximizing the likelihood function, i.e.,
\begin{equation}
\mathcal{L}(A_{0},\alpha,\beta,E_{\rm cut})=\prod_{i}^{n}{\rm Poisson}\left(N_{{\rm obs},i},\;N_{{\rm exp},i}(A_{0},\alpha,\beta,E_{\rm cut})+N_{{\rm bkg},i}\right)\;,
\end{equation}
where $i$ represents the $i$-th energy bin, $N_{{\rm obs},i}$ is the number of observed events from the source,
$N_{{\rm exp},i}$ is the expected signal from the chosen energy spectrum, and $N_{{\rm bkg},i}$ is the estimated
background. For each $E_{\rm cut}$, Ref.~\cite{2021arXiv210612350T} obtained the best-fit spectral parameters
($A_{0}$, $\alpha$, and $\beta$) and the corresponding likelihood value. The statistical significance of the existence
of such a cutoff was calculated using a test statistic (TS) variable, which is the log-likelihood ratio of the fit
with a cutoff and the fit with no cutoff,
\begin{equation}
{\rm TS}(E_{\rm cut})=-\sum_{\rm bin}2\ln\left(\frac{\mathcal{L}_{1}(\hat{\hat{A_{0}}},\hat{\hat{\alpha}},\hat{\hat{\beta}},E_{\rm cut})}{\mathcal{L}_{0}(\hat{A_{0}},\hat{\alpha},\hat{\beta},E_{\rm cut}\rightarrow \infty)}\right)\;,
\end{equation}
where the null hypothesis (without the LIV effect) is the case with $E_{\rm cut}\rightarrow \infty$ and the alternative
hypothesis (with the LIV effect) corresponds to a finite $E_{\rm cut}$. Because the spectral fits do not indicate a
significant preference for $E_{\rm cut}<\infty$, a lower limit on $E_{\rm cut}$ was proceed to set, below which there is weak
or no evidence for the photon decay. This lower limit on $E_{\rm cut}$ also serves as an upper limit on the observed
photon energy $E$. Then Eqs.~(\ref{eq:decay-limit1}), (\ref{eq:decay-limit2}), and (\ref{eq:splitting-limit}) directly
lead to lower limits on $E_{{\rm QG}, 1}$ and $E_{{\rm QG}, 2}$.

\begin{table*}
\caption{Summary of 95\% confidence-level lower limits on the cutoff energy $E_{\rm cut}$ and the linear and quadratic energy scales
of the superluminal LIV as obtained by photon decay ($e^{+}e^{-}$) and photon splitting ($3\gamma$).}
\begin{tabular}{llcccccc}
\hline
\hline
Experiment  & Source & Distance & $E_{\rm cut}$ & $E_{{\rm QG}, 1}$ ${(e^{+}e^{-})}$ &  $E_{{\rm QG}, 2}$ ${(e^{+}e^{-})}$   & $E_{{\rm QG}, 2}$ ${(3\gamma)}$ & Refs. \\
            &        &   [kpc]  &    [TeV]      &      [GeV]        &         [GeV]         &         [GeV]    &       \\
\hline
\multirow{2}{*}{HEGRA}    & Crab Nebula &  2  &   56  & $1.5\times10^{20}$ & $2.8\times10^{12}$ & --- & \cite{2017PhRvD..95f3001M}\\
         & Crab Nebula &  2  &   75  & --- & --- & $1.3\times10^{14}$ & \cite{2019JCAP...04..054A}\\
\hline
Tibet         & Crab Nebula &  2  &   140  & --- & --- & $4.1\times10^{14}$ & \cite{2019EPJC...79.1011S}\\
\hline
\multirow{4}{*}{HAWC}    & J1825--134  & 1.55  &   244  & $1.4\times10^{22}$ & $5.8\times10^{13}$ & $1.2\times10^{15}$ &  \multirow{4}{*}{\cite{2020PhRvL.124m1101A}}\\
    & J1907+063 & 2.37  &   218  & $1.0\times10^{22}$ & $4.7\times10^{13}$ & $1.0\times10^{15}$ &  \\
    & Crab Nebula  & 2  &   152  & $3.4\times10^{21}$ & $2.3\times10^{13}$ & $5.0\times10^{14}$ &  \\
   & J2019+368  & 1.8  &   120  & $1.7\times10^{21}$ & $1.4\times10^{13}$ & $3.2\times10^{14}$ &  \\
\hline
\multirow{5}{*}{LHAASO}    & J2226+6057  & $0.8^{a}$  &   280.7  & $2.1\times10^{22}$ & $7.7\times10^{13}$ & $1.5\times10^{15}$ &  \cite{2021arXiv210507927C}\\
    & J1908+0621  & 2.37 &   370.5  & $4.9\times10^{22}$ & $1.3\times10^{14}$ & $2.8\times10^{15}$ &  \cite{2021arXiv210507927C}\\
     & J1825--1326  & 1.55  &   169.9  & $4.7\times10^{21}$ & $2.8\times10^{13}$ & $0.6\times10^{15}$ &  \cite{2021arXiv210507927C}\\
     & Crab Nebula  & 2 &   750  & $4.0\times10^{23}$ & $5.5\times10^{14}$ & $1.0\times10^{16}$ &  \cite{2021arXiv210612350T}\\
     & J2032+4102  & 1.4  &   1140  & $1.4\times10^{24}$ & $1.3\times10^{15}$ & $2.2\times10^{16}$ &  \cite{2021arXiv210612350T}\\
\hline
\end{tabular}
\label{table3}
\\
$^{a}$The distance of possible celestial objects associated with J2226+6057 was given.
\end{table*}

Table~\ref{table3} lists a summary of 95\% confidence-level lower limits on $E_{\rm cut}$ and the inferred LIV energy scales
obtained through photon decay and photon splitting. As illustrated in Table~\ref{table3}, the highest-energy source LHAASO
J2032+4102 provides the strongest constraints on the energy scales of the superluminal LIV among experimental results with
the same method. The linear LIV energy scale is constrained to be higher than about $10^{5}$ times the Planck energy scale
$E_{\rm Pl}$, and the quadratic LIV energy scales is over $10^{-3}E_{\rm Pl}$.

\subsection{Comparison with different methods}
\begin{figure}[htbp]
\centering
\vskip-0.2in
\hspace*{-45pt}
\includegraphics[width=3.4\textwidth]{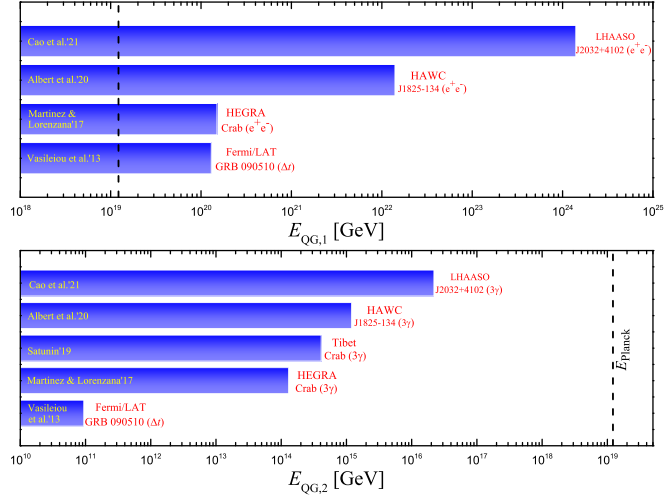}
\vskip-2.7in
\caption{Comparison of the constraints on the superluminal LIV energy scales derived from different experiments
with different methods. We show strong constraints due to photon decay ($e^{+}e^{-}$), photon splitting ($3\gamma$),
and energy-dependent time delay ($\Delta t$). The vertical dashed lines correspond to the Planck energy scale
$E_{\rm Pl}$.}
\label{f4}
\end{figure}

The comparison with the resulting constraints on the superluminal LIV energy scales obtained from different experiments
is shown in Figure~\ref{f4}. We show strong limits on photon decay and photon splitting using ultra-high-energy
gamma-ray observations from HEGRA \cite{2017PhRvD..95f3001M,2019JCAP...04..054A}, Tibet \cite{2019EPJC...79.1011S},
HAWC \cite{2020PhRvL.124m1101A}, and LHAASO \cite{2021arXiv210612350T}. We also show limits due to LIV energy-dependent
time delay searches with the Fermi/LAT \cite{2013PhRvD..87l2001V}. Obviously, the limits on $E_{{\rm QG}, 1}$ and
$E_{{\rm QG}, 2}$ from photon decay and photon splitting are at least four orders of magnitude stronger than those
from the time delay method.

Additionally, observations of linear polarization from astrophysical sources have been widely used to place limits on
the vacuum birefringent parameter $\eta$. By comparing Eqs.~(\ref{eq:LIVdispersion}) and (\ref{eq:dispersion}),
we can derive the conversion from $\eta$ to the limit on the linear LIV energy scale, i.e., $E_{{\rm QG}, 1}=\frac{E_{\rm pl}}{2\eta}$.
At this point, it is interesting to make a comparison of recent achievements in sensitivity of polarization measurements
versus photon decay measurements. The detection of linear polarization in the prompt gamma-ray emission of GRB 061122
yielded the strictest limit on the birefringent parameter, $\eta<0.5\times10^{-16}$, which corresponds to
$E_{{\rm QG}, 1}>1.2\times10^{35}~\mathrm{GeV}$ \cite{2016MNRAS.463..375L,2019MNRAS.485.2401W}. Obviously,
this polarization constraint is about 11 orders of magnitude stronger than the constraint from photon decay
($E_{{\rm QG}, 1}>1.4\times10^{24}~\mathrm{GeV}$) \cite{2021arXiv210612350T}. However, both time-of-flight
constraints and photon decay constraints are essential in an extensive search for nonbirefringent Lorentz-violating
effects.

\section{Summary and outlook}

Many tight constraints on violations of Lorentz invariance have been achieved by observing tiny changes in light that
has propagated over astrophysical distances. Some of these search for a frequency-dependent velocity resulting from
vacuum dispersion. Some works seek a change in polarization arising from vacuum birefringence. Others look for a
hard cutoff in the gamma-ray spectra due to photon decay and photon splitting.

For vacuum dispersion studies, in order to tightly constrain the LIV effects one should choose those explosive
astrophysical sources with shorter spectral lags, higher energy emissions, and longer propagation distances.
As the most energetic bursting events occurring at cosmological sources, GRBs are excellent astrophysical probes
for LIV constraints in the dispersive photon sector. In the past, emission from GRBs had been detected only
at energies below 100 GeV, and LIV limits from time-of-flight measurements of GRBs had also been restricted
in the relatively low energy range. Now that photons of energies above 100 GeV have been detected from two
bright GRBs (GRB 190114C \cite{2019Natur.575..455M} and GRB 180720B \cite{2019Natur.575..464A}), opening
a new spectral window in GRB research. It is expected that such detections will become routine in the future
\cite{2019Natur.575..448Z}, especially with the operations of facilities such as HAWC, LHAASO, and the future
international Cherenkov Telescope Array. Vacuum dispersion constraints will greatly benefit from the observations
of extremely high-energy emission of more GRBs.
For vacuum birefringence studies, in order to obtain tighter constraints on the LIV effects one should choose
those astrophysical sources with longer distances and polarization measurements at higher energies. In this respect,
the ideal sources to seek LIV-induced vacuum birefringence would be polarized gamma-rays from bright GRBs at deep
cosmological distances. The technology for measuring polarization in the 5 to 100 MeV energy range has been constantly
improved and developed. As more and more gamma-ray polarimeters (such as POLAR-II, TSUBAME, NCT/COSI, GRAPE;
see \cite{2017NewAR..76....1M} for a review) enter service, more polarized astrophysical sources at higher
gamma-ray energies are expected to be detected. Polarization measurements of cosmological sources such as GRBs
at such energies will further improve LIV constraints through the vacuum birefringent effect.
For photon decay studies, in order to obtain more stringent bounds on the LIV effects one should choose those
astrophysical sources with abundant ultra-high-energy photons. High energy resolution of spectrum measurements
are desired to determine whether there is a hard cutoff consistent with the prediction of photon decay in the observed spectra
of astrophysical sources. When using ultra-high-energy photons for LIV limits, we should be careful with the uncertainty
for failure to reject a cosmic-ray event. Owing to its unprecedented capability of cosmic ray background rejection
and high energy resolution \cite{2021Natur.594...33C}, LHAASO has competitive advantage in constraining LIV-induced photon decay.


\end{document}